\documentclass[prx,twocolumn,english,superscriptaddress,floatfix,longbibliography]{revtex4-2}

\usepackage{graphicx} 

\usepackage{subcaption}
\usepackage{amsmath}
\usepackage{amssymb}
\usepackage{amsfonts}
\usepackage{dsfont}
\usepackage{dcolumn}
\usepackage{bbm}
\usepackage{ragged2e}
\usepackage{mathtools}
\usepackage{braket}
\usepackage{physics}
\usepackage{siunitx}
\usepackage[unicode]{hyperref}
\usepackage[nameinlink]{cleveref}
\Crefname{section}{Sec.}{Secs.}
\Crefname{equation}{Eq.}{Eqs.}
\Crefname{figure}{Fig.}{Figs.}
\Crefname{tabular}{Tab.}{Tabs.}

\usepackage{bbold} 
\usepackage{nicefrac}
\usepackage{tikz}
\usetikzlibrary{arrows.meta}
\usepackage{lipsum}  
\usepackage{xcolor}

\usepackage[makeroom]{cancel}

\usepackage{xfrac}
\usepackage{svg}

\usepackage[linesnumbered,ruled,vlined]{algorithm2e}
\usepackage{orcidlink}
\usepackage[normalem]{ulem}
\definecolor{tdo_blue}{HTML}{0000FF}
\definecolor{tdo_darkgreen}{HTML}{839A00}
\definecolor{tdo_red}{HTML}{FF0000}

\newcommand{\bes}{\begin{subequations}}
\newcommand{\ees}{\end{subequations}}
\newcommand{\be}{\begin{equation}}
\newcommand{\ee}{\end{equation}}

\makeatletter
  \newcommand\flausr{\@fleqntrue}
\makeatother

\begin{document}

    \title{Variational Corrections to Strong-Disorder Renormalization in the Two-Dimensional Random Transverse Field Ising Model on Quantum Hardware}


    \author{Alina Joch\,\orcidlink{0009-0001-9391-6653}}
    \affiliation{Condensed Matter Theory, TU Dortmund University, Otto-Hahn-Stra\ss{}e 4, 44227 Dortmund, Germany}
    \affiliation{Institute of Software Technology, German Aerospace Center (DLR), 51147 Cologne, Germany}
    
    \author{Benedikt Fauseweh\,\orcidlink{0000-0002-4861-7101}}
    \email[Corresponding author: ]{benedikt.fauseweh@tu-dortmund.de}
    \affiliation{Condensed Matter Theory, TU Dortmund University, Otto-Hahn-Stra\ss{}e 4, 44227 Dortmund, Germany}
    \affiliation{Institute of Software Technology, German Aerospace Center (DLR), 51147 Cologne, Germany}

    \date{\textrm{\today}}

\begin{abstract}
Strong-disorder renormalization-group (SDRG) methods provide an effective description of disordered quantum many-body systems by identifying the dominant couplings for the low energy physics. However, the resulting renormalization-group flow is approximate and neglects correlations beyond the RG description. Here, we investigate whether variational quantum optimization can be used to correct these approximations. Building on the emergent-coupling-based ansatz (ECBA) developed for the one-dimensional random Heisenberg model, we introduce a formulation adapted to the two-dimensional random transverse field Ising model. The SDRG flow determines the structure of the circuit, while variational optimization allows the state to deviate from the RG approximation. We study the resulting ansatz numerically and experimentally on an IQM superconducting quantum processor. For a 16-qubit system, we perform full variational optimization on the quantum processor, while a 49-qubit system is investigated using pre-optimized parameters, achieving a relative energy accuracy of 89\% and 97\%, respectively. Our results demonstrate that RG information can be translated into shallow, hardware-compatible quantum circuits while variational optimization enables corrections beyond the underlying RG approximation also for two-dimensional systems.
\end{abstract}

\maketitle

\section{Introduction}

Disordered quantum many-body systems present a rich variety of low energy phenomena arising from spatial fluctuations in the local couplings and energy scales. In particular, disorder can lead to rare-region effects and unconventional critical behavior associated with infinite-randomness fixed points (IRFPs) \cite{Vojta_2006,vojta19,fisher92}. A powerful framework for studying such systems is the strong-disorder renormalization group (SDRG), which successively eliminates the highest energy degrees of freedom and generates effective interactions between the remaining ones \cite{PhysRevLett.43.1434,PhysRevB.22.1305,igloi05}. By separating the relevant energy scales step by step, SDRG provides the emergent low energy degrees of freedom and has proven highly successful in describing the universal properties of strongly disordered systems. Extensions to excitations \cite{PhysRevX.4.011052} and quantum dynamics \cite{PhysRevLett.110.067204} exist as well. 

The RG flow, however, does not in general represent an exact representation of the underlying many-body problem. The decimation procedure involves approximations to the effective Hamiltonian and the resulting renormalized state, such that information about correlations and quantum fluctuations are discarded during the flow. While these approximations may become asymptotically exact in certain strong-disorder limits \cite{fisher94,fisher95,PhysRevLett.48.344,PhysRevLett.78.1783,PhysRevLett.84.3434,PhysRevLett.89.277203,PhysRevLett.99.230601,PhysRevLett.100.240601,refae04,igloi05}, finite systems and finite disorder strengths can retain contributions that are not captured by the renormalized description. Consequently, the state obtained from an RG construction is understood as a physically motivated approximation to the true ground state rather than an exact representation. An important question is therefore whether the information generated by the RG flow can be used as a starting point for a more accurate description that uses the essential renormalization structure while allowing deviations from the RG approximation to be optimized.

Variational quantum algorithms provide a natural framework for addressing this question. In the variational quantum eigensolver (VQE), a parameterized quantum circuit is used to minimize the expectation value of a target Hamiltonian \cite{peruz14,cerez21,PhysRevResearch.6.043254}, see also Appendix~\ref{app:vqe}. Rather than defining the many-body state completely, the variational parameters allow the circuit to explore a set of quantum states and overcome limitations of an approximate state. The success of this approach depends critically on the choice of ansatz \cite{TILLY20221}. Generic hardware-efficient circuits can provide substantial variational freedom, but may require large circuit depths and can become difficult to optimize as the system size increases \cite{Kandala2017,McClean2018,Cerezo2021}. For disordered systems, where correlations can vary strongly across space, a physically informed circuit layout can provide a more efficient starting point \cite{Watanabe2024VQE}.

\begin{figure}[t] 
    \includegraphics[width=0.8\linewidth]{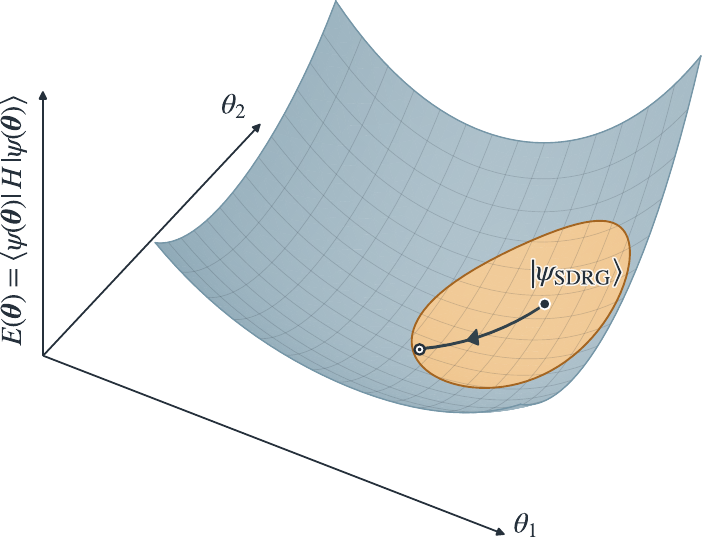}
    \vspace{-0.5em}
    \caption[]{\justifying Schematic illustration of the ECBA. The SDRG state $|\psi_{\mathrm{SDRG}}\rangle$ provides an asymptotically correct starting point on the energy landscape $E(\boldsymbol{\theta})$. The highlighted region represents the variational neighborhood, and the arrow illustrates optimization towards the energy minimum.}
    \label{fig:ecba_landscape}
\end{figure}

This motivates the combination of renormalization-group flow with variational quantum simulation. Instead of interpreting the RG flow as the final description of the many-body state, its output can be used to determine the dominant structure of a variational circuit. The subsequent variational optimization can then modify the parameters of this RG-informed state and thereby account for correlations and energy contributions neglected by the approximate renormalization procedure. In this picture, the RG flow provides the asymptotically correct structure, while the variational optimization provides a mechanism for systematically correcting the resulting approximation, see Fig.~\ref{fig:ecba_landscape}. Related ideas have been developed in tensor-network approaches, where SDRG determines the network geometry and subsequent variational optimization improves the ground-state approximation
\cite{Goldsborough2014,Goldsborough2017,Watanabe2024Structure}.

In our previous work, we introduced the emergent-coupling-based ansatz (ECBA), which follows this principle by using the effective couplings generated during a SDRG procedure to construct a parameterized quantum circuit \cite{Joch_2025, joch2026}. Applied to the one-dimensional random Heisenberg model, the approach demonstrated that the hierarchical structure obtained from the RG flow can be translated into a shallow variational circuit and systematically improve the obtained solution. 

The important next step is to investigate whether this concept remains effective in two dimensions, where the structure of the RG flow becomes more complex. The two-dimensional random transverse-field Ising ferromagnet is particularly suitable for this purpose, due to the presence of an infinite-randomness critical point \cite{pich98,motrunich00,kovac10,Kovac11,kraemer24}. At the same time, the two-dimensional setting requires a reformulation of the ECBA construction. In contrast to the one-dimensional random Heisenberg model the geometry and interaction structure of the two-dimensional Ising model lead to a different pattern of emergent couplings. 

In this work, we develop a two-dimensional formulation of the ECBA formulated for the random transverse field Ising model. We evaluate this approach numerically and experimentally. Numerical simulations are used to investigate the ability of the two-dimensional ECBA to represent the relevant ground states and to study its optimization behavior as the system size is increased. We compare the optimized ECBA state with the corresponding RG approximation and a nearest-neighbor ansatz. This allows us to distinguish the contribution of the renormalization-based initialization from the improvement obtained through variational optimization.
We subsequently implement the complete variational workflow for a 16-qubit system on an IQM superconducting quantum processor, including circuit generation, parameter optimization, and experimental measurement. This allows us to assess not only the expressive power of the proposed ansatz in ideal simulations but also its performance under realistic hardware conditions. We investigate whether the variational optimization systematically lowers the energy relative to the RG approximation and thereby recovers correlations that are not captured by the underlying RG flow. Additionally, we evaluate a 49-qubit system with pre-optimized parameters on the IQM device, addressing the potential for scalability to larger system sizes.

This paper is organized as follows. 
Section \ref{chap:model} presents the two-dimensional random transverse field Ising model, the SDRG procedure used to identify the relevant low energy degrees of freedom, and the construction of the two-dimensional ECBA. 
In Sec.~\ref{chap:results}, we benchmark the ansatz using numerical simulations and compare it with a nearest-neighbor ansatz. Then, it presents the experimental results. Finally, Sec.~\ref{chap:conclusion} summarizes our findings and discusses perspectives for extending renormalization-inspired ansätze.

\section{Model, Renormalization Flow and Quantum Circuit} 
\label{chap:model}

We consider the two-dimensional random transverse field Ising model on a square lattice, described by the Hamiltonian
\begin{equation}
    H = - \sum_{\langle ij \rangle} J_{ij} \sigma_i^z \sigma_j^z - \sum_i h_i \sigma_i^x \, .
\end{equation}
Here, $i$ and $j$ label lattice sites, $\sigma_i^{x,z}$ are the Pauli matrices at site $i$, and the first sum runs over nearest neighbor pairs.
The ferromagnetic exchange couplings $J_{ij} > 0$ and transverse magnetic fields $h_i > 0$ are drawn independently from the distribution \cite{laflo05}
\begin{equation}\label{eq:distribution}
    P(q) = \frac{1}{\delta} q^{-1+\delta^{-1}}
\end{equation}
with $q \in \{J,h\}$. The parameter $\delta \geq 1$ controls the disorder strength.
This distribution is proposed in Ref.~\cite{laflo05} as a natural choice for modeling strong disorder effects because it approaches the fixed-point distribution of the random couplings at the IRFP in the limit $\delta\to\infty$.

\subsection{Strong disorder renormalization group flow}

\begin{figure}[t] 
    \includegraphics[width=0.8\linewidth]{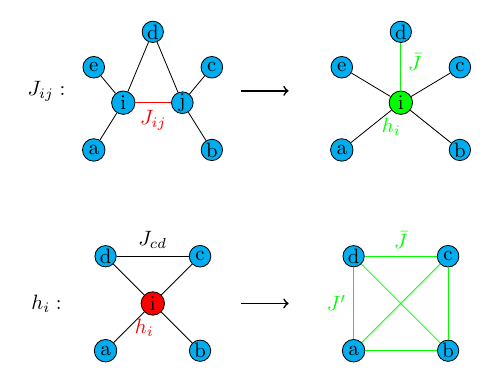}
    \vspace{-0.5em}
    \caption[]{\justifying Illustration of the $J_{ij}$ and the $h_i$ decimation step in the two dimensional SDRG flow \cite{kovac10}. The renormalized parameters are given by $h' \approx h_i h_j/J_{ij}$ and $\bar{J} \approx \max[J_{di},J_{dj}]$ in case of the $J_{ij}$ decimation. For the $h_i$ decimation, they are given by $J'_{ab} \approx J_{ai} J_{bi} / h_i$ and $\bar{J} \approx \max[J_{cd},J'_{cd}]$.}
    \label{fig:sdrg_sketch}
\end{figure}

The strong disorder renormalization group (SDRG) flow introduced by Ma, Dasgupta, and Hu \cite{PhysRevLett.43.1434, kovac10} constructs an effective low energy Hamiltonian. At each step, the term with the largest coupling or magnetic field is decimated, and new terms are generated between the remaining sites using second-order perturbation theory. The procedure is repeated until all terms have been decimated, as illustrated in Fig.~\ref{fig:sdrg_sketch}.

If the largest term is the Ising coupling $J_{ij}$, the two sites $i$ and $j$ connected by this coupling form a GHZ cluster. Consider the Hamiltonian $H_{ij} = -J_{ij} \sigma_i^z \sigma_j^z$. Its ground-state subspace is twofold degenerate and spanned by the states $| \uparrow \uparrow \rangle$ and $|\downarrow \downarrow \rangle$. The degeneracy is lifted by the transverse magnetic fields through second-order perturbation theory. The ground state within this subspace is the GHZ state
\begin{equation}
    |\mathrm{GHZ}\rangle = \frac{1}{\sqrt{2}} (| \uparrow \uparrow \rangle + |\downarrow \downarrow \rangle) \, .
\end{equation}
The two spins can therefore be treated as a single effective spin with the effective magnetic field 
\begin{equation}
    h' \approx \frac{h_i h_j}{J_{ij}} \, .
\end{equation}
In general, arbitrarily large GHZ clusters can form during the RG flow.
If the coupling $J_{ij}$ is part of a triangle, i.e., if sites $i$ and $j$ are both connected to a third site $d$, then the coupling between this third site and the cluster is given by $\bar{J} = \max[J_{di}, J_{dj}]$.

If the largest term is instead the magnetic field $h_i$, site $i$ is eliminated. Effective couplings are generated between all pairs of spins that were connected to site $i$. For two such neighbors, $a$ and $b$, the new coupling is given by
\begin{equation}
    J'_{ab} \approx \frac{J_{ai} J_{bi}}{h_i} \, .
\end{equation}
If site $i$ belongs to a triangle, i.e., if sites $c$ and $d$ are connected both to $i$ and to each other, then the renormalized coupling is given by $\bar{J} = \max[J_{cd},J'_{cd}]$.

Decimating magnetic fields generates new connections between the remaining sites, increasing the number of bonds in the system. This can result in an almost fully connected network, significantly increasing the computational cost of subsequent decimation steps. To avoid this, we use the implementation suggested in Ref.~\cite{kovac10}.
We first identify local maxima, which can be decimated independently. A term is a local maximum if it is at least as large as all neighboring terms. Thus, $J_{ij}$ is a local maximum if $J_{ij} \geq h_i, h_j, J_{ik}, J_{lj} \forall k,l$, and $h_i$ is a local maximum if $h_i \geq J_{ij} \forall j$.
Irrelevant bonds, which never contribute to the renormalization process, can also be filtered out. A bond $J_{ij}$ is irrelevant if sites $i,j$ and $k$ form a triangle in which $J_{ij}$ is the smallest bond and is smaller than the potential new bond $J'_{ij} \approx J_{ik} J_{jk} / h_k$.

The full algorithm begins by identifying all local maxima. Couplings in this list are then decimated in an arbitrary order. After each decimation, the newly generated terms are checked for local maxima. This continues until no local coupling maxima remain. We then select a site whose magnetic field is a local maximum and that has few incident bonds, apply the filtering rule, and decimate the site. All terms are checked again for local maxima, and the procedure is repeated. We refer the reader to Refs.~\cite{igloi05,Igloi2018,lin07,yu08,kovac10} for further details.

\subsection{Identification of the pseudocritical point}

Like the clean transverse field Ising model, the random transverse field Ising model exhibits a quantum critical point separating an ordered and a disordered phase. 
To investigate this transition, we introduce a control parameter $\tau$ that scales the magnetic fields according to $ h_i(\tau) = h_i\exp(\tau)$. Thus, increasing $\tau$ enhances the magnetic fields relative to the couplings.
In two dimensions, the critical point is given by \cite{igloi05, Igloi2018}
\begin{equation}
   \tau_c = \left[ \ln(h) \right]_{av} / \left[ \ln(J) \right]_{av} \, , 
\end{equation}
where $[x]_{av}$ denotes the average of $x$ over quenched disorder.
For $\tau > \tau_c$, the system is in a disordered phase, while for $\tau < \tau_c$, the system is in an ordered phase.
The critical point $\tau_c$ has to be calculated explicitly depending on the disorder distribution. In Ref.~\cite{kovac10}, this was done for a box-like $h$ distribution and a fixed $h$ distribution by using the doubling procedure. 

In the two-dimensional doubling procedure, a square sample is connected to an identical copy at all boundaries. The doubled sample is then renormalized using SDRG, and the resulting cluster structure is analyzed. In particular, we identify clusters that contain sites from both replicas at equivalent positions. These sites are correlated, and the fraction of sites belonging to these clusters defines the replica correlation function. As the control parameter $\tau$ increases, this correlation function gradually decreases until it abruptly drops to zero at a well-defined value, $\tau_c(L)$. We identify $\tau_c(L)$ as the pseudocritical point of the respective sample \cite{kovac10}.

\begin{figure}[t] 
    \includegraphics[width=\linewidth]{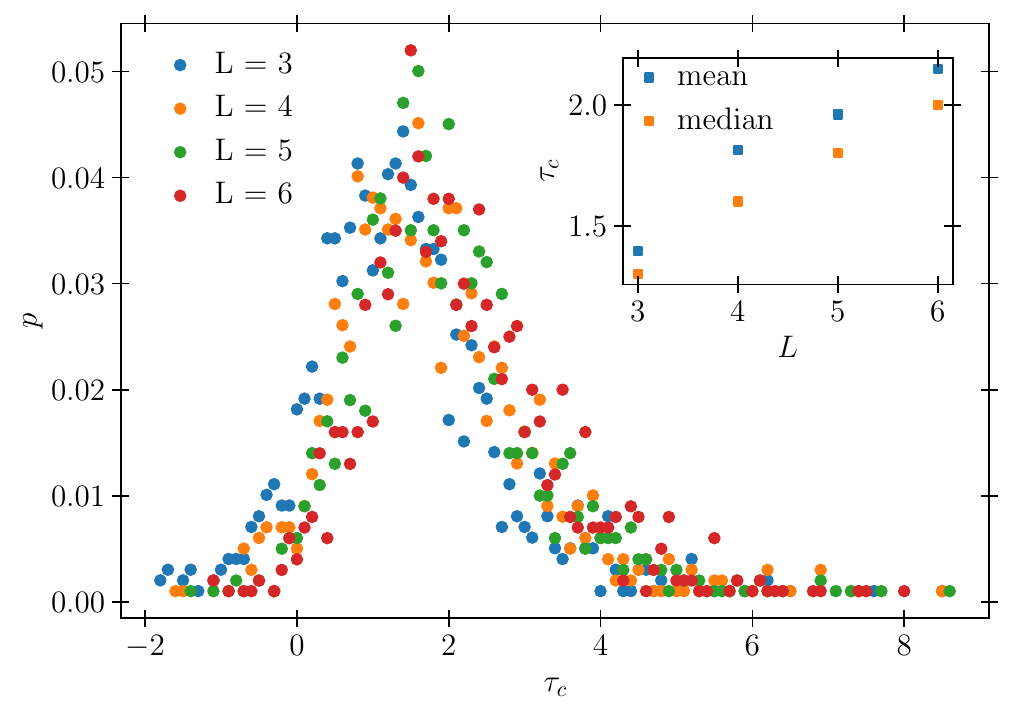}
    \vspace{-0.5em}
    \caption[]{\justifying  Distribution of pseudocritical points for different system sizes at $\delta=2$, obtained from 1000 disorder realizations. The inset shows the mean and median of each distribution.}
    \label{fig:critical_point}
\end{figure}

We investigate both weak and strong disorder by applying the doubling procedure to the distribution in Eq.~\ref{eq:distribution}, with the magnetic fields $h_i$ scaled by $\exp(\tau)$. Figure~\ref{fig:critical_point} shows the resulting distribution of pseudocritical points $\tau_c$, together with its mean and median. We observe a clear dependence on system size. Reference~\cite{kovac10} shows that the critical point converges for sufficiently large systems. In the following simulations and experiments, we choose disorder realizations whose $\tau_c$ is close to the median of the distribution for the corresponding system size.

\subsection{Circuit and embedding}

Our goal is to construct an ansatz that reflects the entanglement structure of the system.
The ansatz consists of an initial state and a variational circuit. We use the state obtained from the RG flow as the initial state.
The variational circuit is designed to correct the RG approximation at steps where the separation of energy scales is weak. Because the generation of effective terms relies on perturbation theory, we identify renormalization steps involving couplings and magnetic fields of similar magnitude.
We first determine all neighboring pairs $(h_i, J_{ij})$ and $(J_{ik}, J_{kj})$ with $J_{ij} = J_{ji}$ and $i \neq j$, for which both values exceed a threshold $s$ and their relative difference is smaller than a tolerance $t$. We then apply $R_{zz}$ entangling gates to the corresponding qubit pairs to introduce variational corrections at RG steps where the perturbative approximation is least reliable.
The $R_{zz}$ gate is parameterized by an angle $\theta_i$ and can realize the identity, providing greater variational flexibility than fixed entangling gates such as the CNOT.
After these entangling gates, we apply an $R_x$ rotation to every qubit. Finally, we apply entangling gates within the clusters obtained from the RG flow, followed by another rotation on each qubit belonging to these clusters. Following the terminology of Ref.~\cite{joch2026}, we call this the emergent-coupling-based ansatz (ECBA). The circuit is illustrated in Fig.~\ref{fig:rg_ansatz}, and pseudocode is provided in Appendix~\ref{app:pesudocode}.

\begin{figure}[t] 
    \includegraphics[width=0.8\linewidth]{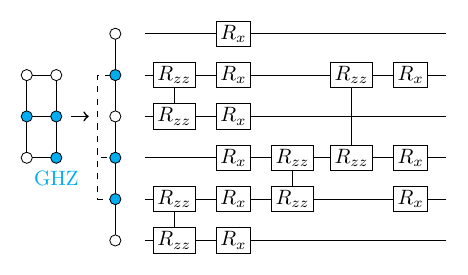}
    \vspace{-0.5em}
    \caption[]{\justifying  Schematic of the ECBA for 6 qubits. The initial state consists of GHZ clusters and qubits aligned along the $x$ direction according to the RG flow. The variational circuit begins with entangling gates associated with terms of similar magnitude, followed by $R_x$ gates on every qubit. Then, $R_{zz}$ and $R_x$ gates are applied to the clusters.}
    \label{fig:rg_ansatz}
\end{figure}

\begin{figure}[t] 
    \includegraphics[width=0.8\linewidth]{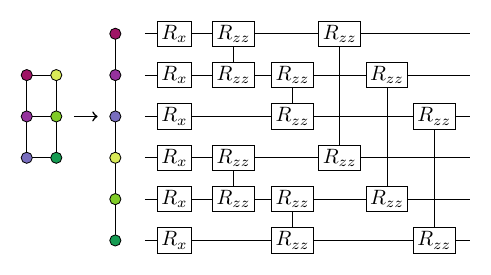}
    \vspace{-0.5em}
    \caption[]{\justifying  Schematic of the NNA for 6 qubits. The variational circuit consists of $R_x$ gates applied to all qubits followed by $R_{zz}$ gates on nearest-neighbor pairs.}
    \label{fig:ansatz_hea}
\end{figure}

We compare the ECBA with a nearest-neighbor ansatz (NNA) consisting of rotation gates on all qubits followed by entangling gates on all nearest-neighbor pairs. The NNA is hardware-efficient by construction. The ECBA can also be hardware-efficient in the sense that gates act only on neighboring qubits. However, the NNA denotes an ansatz that uses the lattice connectivity without incorporating information about the disorder realization or the RG flow. This ansatz is illustrated in Fig.~\ref{fig:ansatz_hea}.   

The threshold $s$ and tolerance $t$ control the construction of the ECBA. We find that a high tolerance improves the results and therefore set $t=0.99$. The threshold $s$ controls the number of pairs in the first part of the circuit. Lowering $s$ improves accuracy but also increases the number of parameters and the circuit depth, making an embedding without SWAP gates more difficult. The threshold must therefore be chosen with the embedding constraints in mind. We generally use $s = 1.0$.
An embedding without SWAP gates also requires careful construction of the GHZ states associated with the clusters. We construct the preparation circuit so that each spin participates in entangling gates with at most three other spins, while keeping the circuit depth as low as possible.

\section{Results}
\label{chap:results}

\subsection{Classical simulations}

\begin{figure}[t] 
    \includegraphics[width=\linewidth]{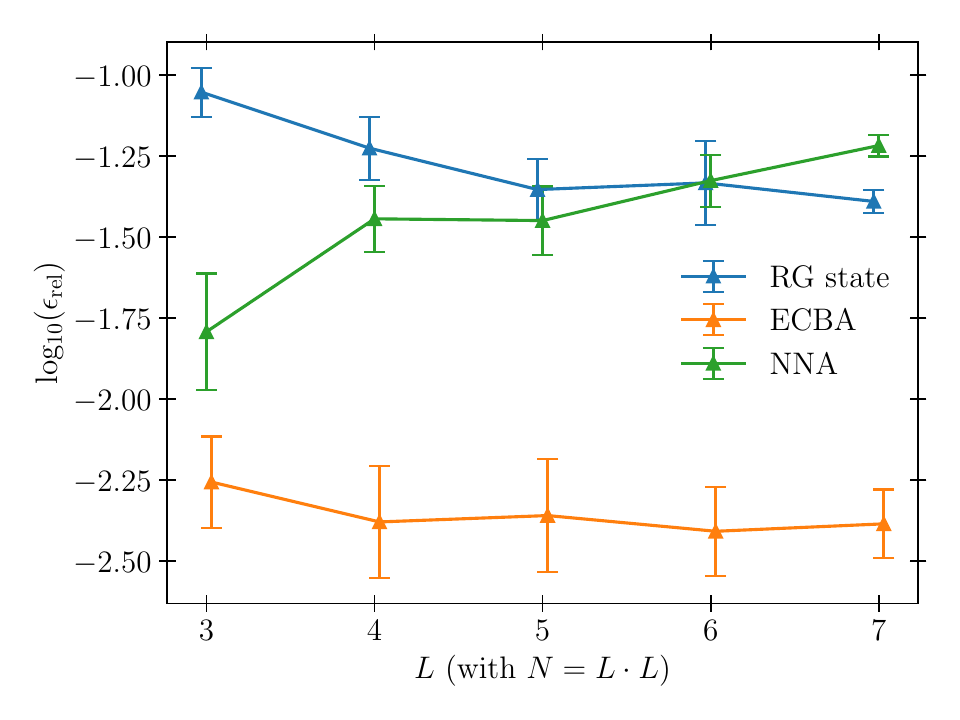}
    \vspace{-0.5em}
    \caption[]{\justifying Logarithm of the relative energy error with respect to the tensor network result as a function of system size for $\delta=2$. Results are averaged over 10 configurations for $L=3,4$ and over 6 configurations for $L=5,6,7$.}
    \label{fig:simulation_L}
\end{figure}

\begin{figure}[t] 
    \includegraphics[width=\linewidth]{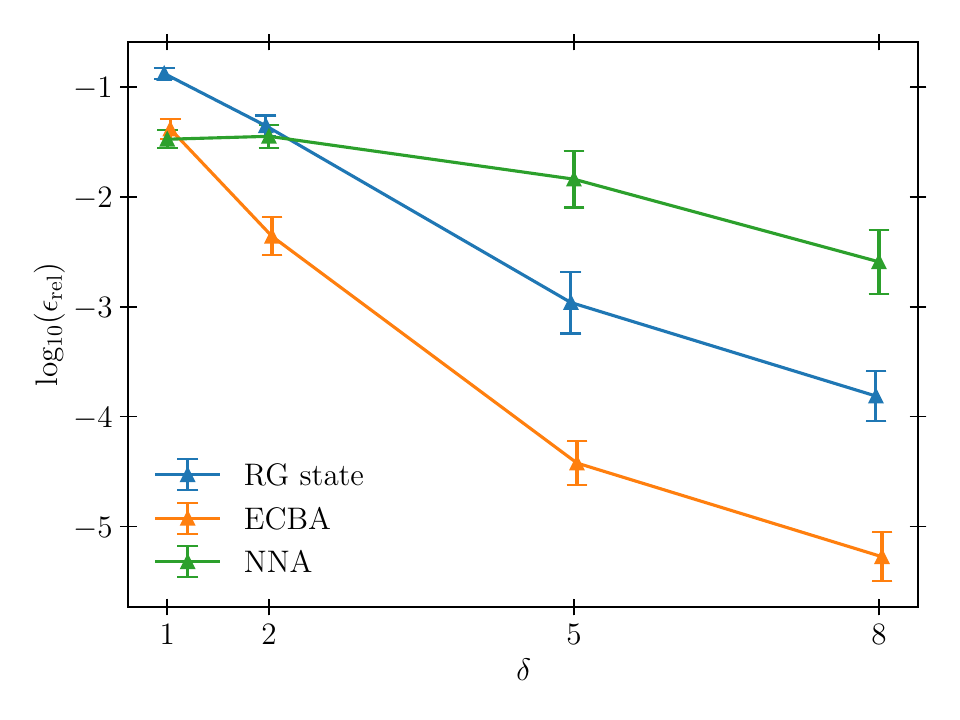}
    \vspace{-0.5em}
    \caption[]{\justifying Logarithm of the relative energy error with respect to the tensor network result as a function of disorder strength $\delta$ for $L=5$. Results are averaged over 6 configurations for each value of $\delta$.}
    \label{fig:simulation_delta}
\end{figure}

First, we analyze the ECBA using classical matrix product state simulations for different system sizes and disorder strengths $\delta$, for details on the numerics see Appendix~\ref{app:numericaldetails}. For each combination of system size and disorder strength, we choose configurations whose pseudocritical points $\tau_c$ are close to the median of the corresponding distribution.
The results are shown in Figs.~\ref{fig:simulation_L} and \ref{fig:simulation_delta}. In Fig.~\ref{fig:simulation_L}, the logarithm of the relative error with respect to a tensor-network-based DMRG calculation $E_\mathrm{tn}$
\begin{equation}
    \epsilon_\mathrm{rel} = \frac{|E_\mathrm{tn}- E_\mathrm{VQE}|}{|E_\mathrm{tn}|} 
\end{equation}
is plotted for different ansätze as a function of the system size $N=L \cdot L$ at $\delta =2$. The RG initial state alone, without a variational circuit, provides a good initial approximation to the ground-state energy. Adding a variational circuit based on the RG structure reduces the relative error by approximately one order of magnitude. The NNA also performs well for smaller systems, but its relative error increases with system size. The ECBA outperforms both the RG initial state and the NNA in all cases considered.

Table~\ref{tab:numberparams} lists the number of variational parameters in the two ansätze for different system sizes. The ECBA and the NNA have comparable parameter counts. For the NNA, the number of parameters is independent of the disorder configuration and equals $3N$. For the ECBA, it depends on the configuration, the threshold $s$, and the tolerance $t$. For $t=0.99$ and $s=1.0$, we observe linear scaling with a smaller slope than for the NNA. 

For the systems studied here, the circuit depths of both the NNA and the ECBA are largely independent of system size. They are primarily determined by the gate ordering and, for the ECBA, by the parameters $s$ and $t$. For the ECBA, the circuit depth across the configurations studied here is $21\pm1$. The key to achieving a low circuit depth is to construct the GHZ circuit efficiently.
For the NNA, the circuit depth depends strongly on the order in which the entangling gates are applied. On average, for the systems considered here, the corresponding circuit depth is $32\pm5$.

In Appendix~\ref{app:gradient}, we investigate the gradient variance for both ansätze as a function of system size. For the system sizes and circuit depths considered here, we observe no significant decrease in the gradient variance with increasing system size. This suggests that barren plateaus do not pose a significant challenge for the ansätze studied in this work.

\begin{table}[htb]
\centering
    \begin{tabular}{l|c c c c c}
        $L$ & 3 \,\,  & 4 \,\,   & 5 \,\,   & 6\,\,   &7 \\ \hline 
        NNA  & 27 & 48 & 75 & 108 & 147 \\
        ECBA  & $24 \pm 4$ & $42 \pm 3$ & $57 \pm 5$ & $88 \pm 9$ & $120 \pm 10$
    \end{tabular}
    \caption{\label{tab:numberparams} \justifying Comparison of number of variational parameters for the NNA and the ECBA for different system sizes $N=L\cdot L$. The values given for the ECBA are mean values for different configurations with $t=0.99$ and $s=1.0$. }
\end{table}

Figure~\ref{fig:simulation_delta} shows the dependence on disorder strength $\delta$ for $L=5$. As the disorder strength increases, the ECBA becomes more accurate, while the performance of the NNA deteriorates relative to both the RG state and the ECBA. 
We attribute this trend to the greater accuracy of the RG approximation for large systems and strong disorder. For large systems with weak disorder, the ECBA is therefore a promising option when the NNA performs poorly and an improvement over the RG approximation is desired.

\subsection{Quantum simulations}

\begin{figure*}[t]
    \vspace{-1.5em}
    \hspace{-1em}
    \begin{subfigure}[t]{0.37\textwidth}
          \includegraphics[width=\linewidth]{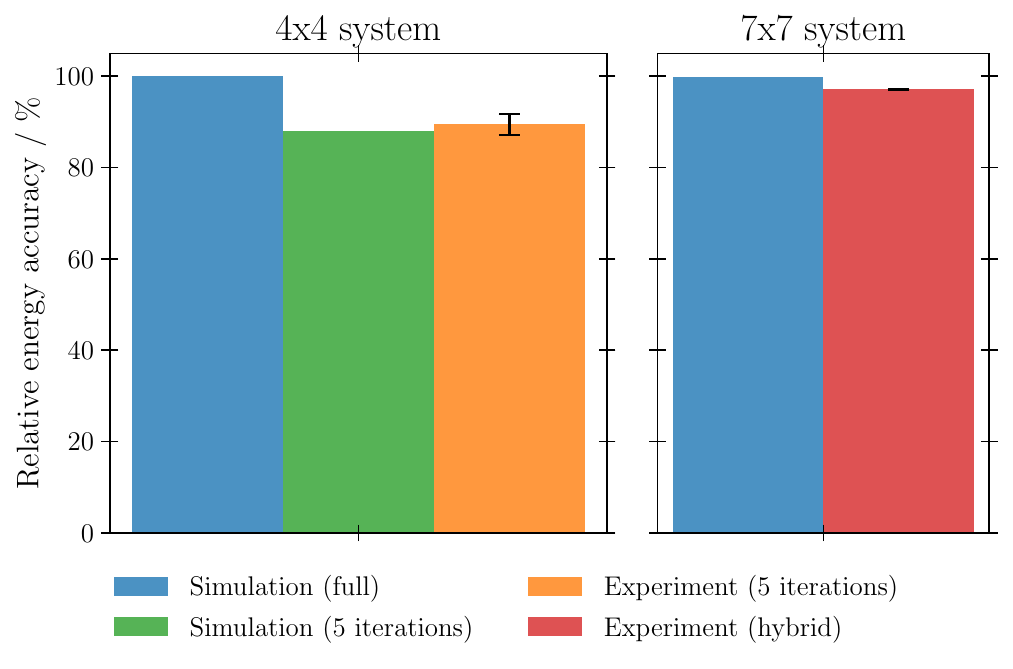}
          \caption[]{}
          \label{fig:energy}
    \end{subfigure}
    \begin{subfigure}[t]{0.25\textwidth}

\begin{tikzpicture}[scale=0.75, transform shape]

\def\abstand{0.9}

\node[draw,circle,inner sep=1.9pt,fill=cyan] (QB1) at (2*\abstand,1*\abstand) {1};
\node[draw,circle,inner sep=1.9pt,fill=cyan] (QB2) at (3*\abstand,1*\abstand) {2};
\draw[color=cyan,ultra thick] (QB1) -- (QB2);

\node[draw,circle,inner sep=1.9pt,fill=cyan] (QB3) at (1*\abstand,2*\abstand) {3};
\node[draw,circle,inner sep=1.9pt,fill=cyan] (QB4) at (2*\abstand,2*\abstand) {4};
\node[draw,circle,inner sep=1.9pt,fill=cyan] (QB5) at (3*\abstand,2*\abstand) {5};
\node[draw,circle,inner sep=1.9pt,fill=cyan] (QB6) at (4*\abstand,2*\abstand) {6};
\node[draw,circle,inner sep=1.9pt,fill=cyan] (QB7) at (5*\abstand,2*\abstand) {7};
\draw[] (QB1) -- (QB4);
\draw[color=cyan,ultra thick] (QB2) -- (QB5);
\draw[color=cyan,ultra thick] (QB3) -- (QB4);
\draw[color=cyan,ultra thick] (QB4) -- (QB5);
\draw[color=cyan,ultra thick] (QB5) -- (QB6);
\draw[] (QB6) -- (QB7);
\draw[color=green, dashed, ultra thick] (QB6) -- (QB7);

\node[draw,circle,inner sep=1.9pt,fill=cyan] (QB8) at (1*\abstand,3*\abstand)  {8};
\node[draw,circle,inner sep=1.9pt,fill=cyan] (QB9) at (2*\abstand,3*\abstand)  {9};
\node[draw,circle,inner sep=0.5pt,fill=cyan] (QB10) at (3*\abstand,3*\abstand) {10};
\node[draw,circle,inner sep=0.5pt,fill=cyan] (QB11) at (4*\abstand,3*\abstand) {11};
\node[draw,circle,inner sep=0.5pt,fill=cyan] (QB12) at (5*\abstand,3*\abstand) {12};
\draw[] (QB3) -- (QB8);
\draw[color=green, dashed, ultra thick] (QB3) -- (QB8);
\draw[] (QB4) -- (QB9);
\draw[color=green, dashed, ultra thick] (QB4) -- (QB9);
\draw[] (QB5) -- (QB10);
\draw[color=cyan,ultra thick] (QB6) -- (QB11);
\draw[color=cyan,ultra thick] (QB7) -- (QB12);
\draw[] (QB8) --  (QB9);
\draw[] (QB9) --  (QB10);
\draw[color=cyan,ultra thick] (QB10) -- (QB11);
\draw[] (QB11) -- (QB12);

\node[draw,circle,inner sep=0.5pt,fill=white] (QB13) at (1*\abstand,4*\abstand) {13};
\node[draw,circle,inner sep=0.5pt,fill=white] (QB14) at (2*\abstand,4*\abstand) {14};
\node[draw,circle,inner sep=0.5pt,fill=white] (QB15) at (3*\abstand,4*\abstand) {15};
\node[draw,circle,inner sep=0.5pt,fill=cyan] (QB16) at (4*\abstand,4*\abstand) {16};
\node[draw,circle,inner sep=0.5pt,fill=cyan] (QB17) at (5*\abstand,4*\abstand) {17};
\draw[] (QB8) --  (QB13);
\draw[] (QB9) --  (QB14);
\draw[] (QB10) -- (QB15);
\draw[color=cyan,ultra thick] (QB11) -- (QB16);
\draw[] (QB12) -- (QB17);
\draw[] (QB13) -- (QB14);
\draw[] (QB14) -- (QB15);
\draw[] (QB15) -- (QB16);
\draw[] (QB16) -- (QB17);
\draw[color=green, dashed, ultra thick] (QB16) -- (QB17);

\node[draw,circle,inner sep=0.5pt,fill=white] (QB18) at (2*\abstand,5*\abstand) {18};
\node[draw,circle,inner sep=0.5pt,fill=cyan] (QB19) at (3*\abstand,5*\abstand) {19};
\node[draw,circle,inner sep=0.5pt,fill=cyan] (QB20) at (4*\abstand,5*\abstand) {20};
\draw[] (QB14) -- (QB18);
\draw[] (QB15) -- (QB19);
\draw[] (QB16) -- (QB20);
\draw[color=green, dashed, ultra thick] (QB16) -- (QB20);
\draw[] (QB18) -- (QB19);
\draw[] (QB19) -- (QB20);
\draw[color=green, dashed, ultra thick] (QB19) -- (QB20);

\def\shift{0.25}
\node[draw=none,circle,inner sep=1.9pt,fill=none, text=red] (SP10) at (3*\abstand+\shift,5*\abstand+\shift) {16};
\node[draw=none,circle,inner sep=1.9pt,fill=none, text=red] (SP11) at (4*\abstand+\shift,5*\abstand+\shift) {13};

\node[draw=none,circle,inner sep=1.9pt,fill=none, text=red] (SP11) at (4*\abstand+\shift,4*\abstand+\shift) {9};
\node[draw=none,circle,inner sep=1.9pt,fill=none, text=red] (SP12) at (5*\abstand+\shift,4*\abstand+\shift) {12};

\node[draw=none,circle,inner sep=1.9pt,fill=none, text=red] (SP8) at (1*\abstand+\shift,3*\abstand+\shift)  {8};
\node[draw=none,circle,inner sep=1.9pt,fill=none, text=red] (SP9) at (2*\abstand+\shift,3*\abstand+\shift)  {2};
\node[draw=none,circle,inner sep=1.9pt,fill=none, text=red] (SP10) at (3*\abstand+\shift,3*\abstand+\shift) {14};
\node[draw=none,circle,inner sep=1.9pt,fill=none, text=red] (SP11) at (4*\abstand+\shift,3*\abstand+\shift) {3};
\node[draw=none,circle,inner sep=1.9pt,fill=none, text=red] (SP12) at (5*\abstand+\shift,3*\abstand+\shift) {10};

\node[draw=none,circle,inner sep=1.9pt,fill=none, text=red] (SP3) at (1*\abstand+\shift,2*\abstand+\shift) {5};
\node[draw=none,circle,inner sep=1.9pt,fill=none, text=red] (SP4) at (2*\abstand+\shift,2*\abstand+\shift) {1};
\node[draw=none,circle,inner sep=1.9pt,fill=none, text=red] (SP5) at (3*\abstand+\shift,2*\abstand+\shift) {11};
\node[draw=none,circle,inner sep=1.9pt,fill=none, text=red] (SP6) at (4*\abstand+\shift,2*\abstand+\shift) {7};
\node[draw=none,circle,inner sep=1.9pt,fill=none, text=red] (SP7) at (5*\abstand+\shift,2*\abstand+\shift) {6};

\node[draw=none,circle,inner sep=1.9pt,fill=none, text=red] (SP4) at (2*\abstand+\shift,1*\abstand+\shift) {15};
\node[draw=none,circle,inner sep=1.9pt,fill=none, text=red] (SP5) at (3*\abstand+\shift,1*\abstand+\shift) {4};

\end{tikzpicture}

          \caption[]{}
          \label{fig:embedding_L4}
    \end{subfigure}
    \hspace{-2em}
    \begin{subfigure}[t]{0.38\textwidth}
          \input{texfigures/embedding_L7.tex}
          \caption[]{}
          \label{fig:embedding_L7}
    \end{subfigure}

     \hspace{-3em}
     \begin{subfigure}[t]{1.05\textwidth} 
          \includegraphics[width=1.\linewidth]{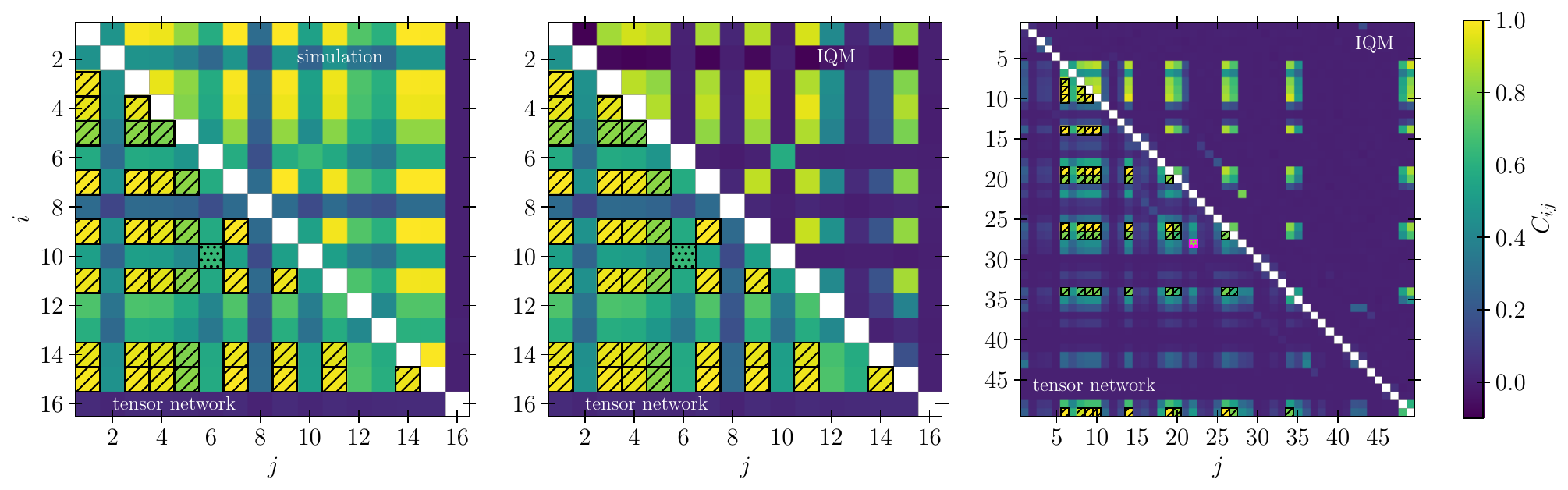}
          \vspace{-2.0em}
          \caption[]{}
          \label{fig:heatmaps}
     \end{subfigure}

     \caption{\justifying Quantum simulation results for the ECBA approach.
                (a) Relative energy accuracy for the $L=4$ and $L=7$ systems at $\delta=2$. For the smaller system, we compare the full VQE simulation with a simulation limited to 5 iterations and an experimental VQE run with 5 iterations. For the larger system, we compare the full VQE simulation with experimental measurements using pre-optimized parameters. The error-mitigation techniques used for both systems are described in Appendix~\ref{app:experimental_details}.
        (b)/(c) Embeddings of the $L=4$ configuration on the IQM Garnet device (b) and the $L=7$ configuration on the IQM Emerald device (c). Green dashed lines indicate pairs selected for variational corrections, blue lines denote gates applied within the RG clusters, and red numbers indicate the mapping of logical qubit indices to physical qubits.
        (d) 
        Heatmaps of $C_{ij}$ for the $L=4$ and $L=7$ systems. 
        For $N=16$ spins, the tensor network values in the lower-left triangles are compared with the simulated VQE results (left) and the experimental IQM results after 5 iterations (center) in the upper-right triangles. The two clusters obtained from the RG flow for the $L=4$ configuration are marked with diagonal hatching and a dotted pattern, respectively.
        For $N=49$ spins (right), the tensor network values in the lower-left triangle are compared with the IQM results obtained using pre-optimized parameters in the upper-right triangle. The two RG clusters for this configuration are marked with diagonal hatching and a magenta dotted pattern, respectively.      }
        \label{fig:results_experiment}

\end{figure*}

We next investigate the performance of the ECBA on quantum hardware. We use a single realization, in which we draw the couplings and fields from a distribution with disorder strength of $\delta=2$. The results are summarized in Fig.~\ref{fig:results_experiment}. We first consider a system of $N=7 \cdot 7$ spins on the IQM Emerald device. We pre-optimize the variational parameters using classical simulations and then estimate the energy on the quantum device by executing the optimized circuit. This yields a relative energy accuracy of $97.09\%\pm0.07\%$. 

Using pre-optimized parameters leaves most of the computation to classical simulations. We therefore also perform an experimental VQE run in which the energy and gradients are evaluated on the quantum device throughout the optimization, while the parameter updates are computed classically. Each optimization step requires the gradient in addition to the energy. For these hardware evaluations, we use the parameter-shift rule, which requires two energy evaluations per variational parameter to compute the gradient. We therefore choose a smaller system of $N = 4\cdot4$ spins and apply error mitigation only during the final optimization iterations and the final energy evaluation.
After 5 iterations on the IQM device, we achieve a relative energy accuracy of $89.38\%\pm2.26\%$. This is close to the simulated accuracy of $87.98\%$ after 5 iterations. The energy results are shown in Fig.~\ref{fig:energy} and the convergence behavior is described in Appendix~\ref{app:convergence}. The hardware embeddings for the $L=4$ and $L=7$ systems are shown in Figs.~\ref{fig:embedding_L4} and \ref{fig:embedding_L7}, respectively.

While the VQE optimizes the variational state by minimizing its energy, the energy alone does not fully characterize the properties of the resulting quantum state. We therefore also examine its correlation structure. The SDRG provides a nontrivial initial state whose GHZ clusters already encode correlations. Near the quantum critical point, these clusters can extend over large spatial distances, reflecting the growth of the correlation length. Thus, the initial state already contains nontrivial long-range correlations, which are further modified by the variational optimization. To investigate these changes, we consider the correlation matrix $C_{ij} = \langle Z_iZ_j \rangle$. 

We first consider the $N = 4\cdot4$ system. Figure~\ref{fig:heatmaps} (left) compares the correlation matrices obtained from the ECBA simulation and the tensor network calculation. The simulated data closely match the tensor network values. The two clusters in this configuration are marked with diagonal hatching and a dotted pattern, respectively. The off-diagonal correlation structure follows these clusters. The simulation accurately reproduces both short- and long-range correlations. 
Figure~\ref{fig:heatmaps} (center) compares the experimental correlation matrix after 5 ECBA iterations with the tensor network results. The experimental results are less accurate than the simulations. Correlations at intermediate distances are missing, which may partly reflect the limited number of iterations. Nevertheless, the overall correlation structure is already clearly recognizable.

Figure~\ref{fig:heatmaps} (right) compares the tensor network correlation matrix $C_{ij} = \langle Z_iZ_j \rangle$ for the $N = 7\cdot7$ system with the results obtained on the IQM device using pre-optimized parameters. 
The tensor network heatmap reveals a richer structure, with a larger number of values deviating from zero. The experimental results reproduce the structures surrounding the RG clusters, capturing both short- and long-range correlations with high accuracy. 
Unlike in the smaller system, some groups of spins are not connected to one another by gates. Consequently, correlations between these spins are not captured in the IQM results. For example, the embedding in Fig.~\ref{fig:embedding_L7} shows that spin 7 is connected by gates to the large cluster including the spins [49, 6, 20, 19, 26, 27, 8, 9, 14, 20, 34]. The corresponding correlations agree well with the tensor network results. However, spin 7 is not connected to spins such as 42 and 43, for which the tensor network calculation shows clearly nonzero correlations that are absent in the IQM results. To capture these additional correlations, the values of $s$ and $t$ would need to be adjusted, at the cost of introducing more parameters and potentially increasing the circuit depth.

\section{Conclusions}
\label{chap:conclusion}

In this work, we investigate how SDRG information can be incorporated into variational quantum circuits for disordered quantum many-body systems. Building on the emergent-coupling-based ansatz (ECBA), originally developed for the one-dimensional random Heisenberg model \cite{joch2026}, we develop a formulation adapted to the two-dimensional random transverse field Ising model. The SDRG flow provides a physically motivated circuit structure based on the dominant effective couplings, while variational optimization allows the quantum state to deviate from the approximate RG description and capture correlations beyond those encoded by the renormalization procedure. This enables variational corrections to the RG approximation and improves the resulting ground-state description.

We study this ansatz using numerical simulations and experiments on IQM devices. 
The simulations show that the ECBA outperforms both the classical RG approximation and a nearest-neighbor ansatz across the system sizes and disorder strengths considered Only when approach the clean limit the ECBA performs similar to the NNA.
For a 16-qubit system, we evaluate energies and gradients on quantum hardware throughout 5 variational optimization iterations, achieving a relative energy accuracy of 89\%. For a 49-qubit system, we use pre-optimized parameters and achieve a relative energy accuracy of 97\%. These results demonstrate that RG information can guide the construction of efficient variational quantum circuits while retaining the flexibility to improve upon the underlying RG approximation. The experiments establish the feasibility of this strategy on state-of-the-art superconducting quantum hardware.


Our results show that the ECBA approach applies to a broad class of disordered quantum system which can be tackled using SDRG. The successful application to two-dimensional systems demonstrates the potential of this approach for studying quantum many-body systems with more complex connectivity and correlation structures, for which accurate numerical simulations remain challenging \cite{Chen2024, Fauseweh2023quantumcomputing, fause24}. Future work could investigate the performance of this strategy for larger systems, alternative disorder distributions \cite{Igloi2018}, and other quantum many-body models, such as the Ashkin-Teller model \cite{Igloi2018} and the random Hubbard model \cite{igloi06}. 

An additional promising direction is the application of the ECBA ansatz to molecular electronic-structure problems \cite{li2019, yang2024}. Accurately describing electronic correlations in molecular systems remains a major challenge in quantum chemistry, where the VQE is commonly employed with hardware-efficient ansätze or chemically motivated approaches such as unitary coupled cluster with singles and doubles (UCCSD) \cite{omall16, Kandala2017, Masso15, Lolur2023, naseri25, bistafa25}. Investigating whether the ECBA approach can be adapted to molecular Hamiltonians could therefore provide an alternative strategy for constructing physically motivated ansätze. In particular, the results for two-dimensional systems provide a useful stepping stone toward exploring more complex interaction structures.

Finally, the effective Floquet Hamiltonian offers a potential way for extending the method to non-equilibrium quantum dynamics \cite{Eisert2015,Fauseweh2021,PhysRevResearch.6.033092}, particularly in systems exhibiting dimerized structures \cite{Katz2025}. Relevant examples include one-dimensional periodically driven Floquet systems governed by infinite-randomness fixed points under SDRG flow \cite{Berda2018}, disorder-induced critical phases in periodically driven dimerized Kitaev chains \cite{Roy2024}, and symmetry-enriched random quantum spin chains \cite{duque21}. Exploring these systems could help establish whether the ECBA framework can also capture the effective correlations arising in periodically driven disordered quantum systems.

\begin{acknowledgments} 
This work has been supported by the Quantum Fellowship Program of the German Aerospace Center (DLR). This work was funded by the Deutsche Forschungsgemeinschaft (DFG, German Research Foundation)—Project No. FA 1884/5-1. We acknowledge IQM for providing access to their quantum computing resources and for their technical support. We thank Yanick Kind and Kevin Lively for helpful discussions and support with the numerical implementation and code development. We acknowledge the use of AI tools for language editing and programming assistance.
\end{acknowledgments}

\bibliography{sample.bib}

\onecolumngrid

\begin{appendix}
\section{VQE}
\label{app:vqe}
The Variational Quantum Eigensolver (VQE) is a hybrid quantum-classical method used to obtain an approximate ground state of a Hamiltonian $H$ describing a system of $n$ qubits. The algorithm employs a parameterized quantum circuit represented by the unitary operator $U(\theta)$, where the parameters $\theta$ are adjusted to find a quantum state $U(\theta) |0\rangle^{\otimes n}$ that approximates the ground state as closely as possible.

The performance of the variational ansatz is characterized through the expectation value of the Hamiltonian
\begin{equation}
    E(\boldsymbol{\theta}) = \langle \psi(\boldsymbol{\theta}) | H | \psi(\boldsymbol{\theta}) \rangle \, .
\end{equation}
This function serves as the objective function of the VQE algorithm. Its value is estimated on a quantum device, while a classical optimization procedure iteratively modifies the circuit parameters to minimize the energy. After convergence, the optimized parameters $\boldsymbol{\theta}_{\mathrm{opt}}$ determine the approximate ground state,
\begin{equation}
    |\psi(\boldsymbol{\theta}_{\mathrm{opt}})\rangle = U(\boldsymbol{\theta}_{\mathrm{opt}}) |0\rangle^{\otimes n} \, ,
\end{equation}
with the associated energy $E(\boldsymbol{\theta}_{\mathrm{opt}})$. According to the variational principle, this energy constitutes an upper bound on the exact ground-state energy.

\section{RG-based selection of variational couplings}
\label{app:pesudocode}
The construction of the variational circuit follows a two-stage procedure. First, clusters identified through the RG-based selection algorithm are incorporated into the circuit. These clusters are determined using the algorithm presented in Ref.~\cite{kovac10}. Second, neighboring couplings $J_{ij}$ and local fields $h_i$ of comparable magnitude are identified. These matching terms are then used to further refine the circuit structure. The procedure for identifying these terms, together with the subsequent construction of the variational circuit, is shown in the following. 

\begin{algorithm}[H]
\SetAlgoLined
\KwIn{Local fields $h_i$, couplings $J_{ij}$, tolerance $t$, threshold $s$}
\KwOut{Matched bonds $M$}

Construct neighbor sets $N(i)$ from $J$\;
Initialize matches $M$, used sites $S\gets\emptyset$\;
$P\gets\{(h,i,h_i):|h_i|>s\}\cup
\{(J,(i,j),J_{ij}):|J_{ij}|>s\}$\;
Sort $P$ by decreasing value $|v|$\;

\ForEach{$(\mathrm{type \,} q, \mathrm{index \,} x, \mathrm{value \,} v)\in P$}{
    \eIf{$q=h$}{
        $i\gets x$\;
        \If{$i\notin S$}{
            $b^*\gets\arg\min_{b=(i,j)\in J}\mathrm{rel\_diff}(v,J_b)$, \;
            $e^*\gets\mathrm{rel\_diff}(v,J_{b^*})$\;
            \If{$b^*\neq\mathrm{None \, and\,} e^*<t$ $\mathrm{and}$ $b^*\notin M$}{
                $M,S\gets M\cup\{b^*\},S\cup\{i\}$\;
            }
        }
    }{
        $b\gets\mathrm{pair}(i,j)$, where $x=(i,j)$\;
        \If{$b\notin M$}{
            $C\gets\{h_i,h_j\}\cup
            \{J_{k\ell}: k\in\{i,j\},\ \ell\in N(k),
            \ell\notin\{i,j\},\ (k,\ell)\in J\}$\;
            $e^*\gets\min_{c\in C}\mathrm{rel\_diff}(v,c)$\;
            \If{$e^*<t$}{
                $M\gets M\cup\{b\}$\;
            }
        }
    }
}
\Return{$M$}\;
\caption{Identification of matching fields and couplings}
\end{algorithm}

\begin{algorithm}[H]
\SetAlgoLined
\KwIn{Model $H$, clusters $C$, connected clusters $C_{\mathrm{conn}}$, tolerance $t$, selection threshold $s$}
\KwOut{Variational circuit $\mathcal{C}_{\mathrm{RG}}$, number of parameters $N_{\mathrm{par}}$}

Determine matching pairs $M$ from the model using $t$ and $s$\;
Initialize empty gate list $\mathcal{G}$ and parameter counter $p \gets 1$\;

\ForEach{ $\mathrm{pair \,} (i,j) \in M$}{
    $\mathrm{Append \, } R_{zz}(i,j,p) \mathrm{\, to \, } \mathcal{G}$\;
    $p \gets p+1$\;
}

\ForEach{ $\mathrm{site \, } i \in \mathrm{\, model}$}{
    $\mathrm{Append \, } R_x(i,p) \mathrm{\, to \, } \mathcal{G}$\;
    $p \gets p+1$\;
}

\ForEach{$c \in C$}{
    \ForEach{$(i,j) \mathrm{\, of\, consecutive \, sites} \in c$}{
        Append $R_{zz}(i,j,p)$ to $\mathcal{G}$\;
        $p \gets p+1$\;
    }
}

\ForEach{$c \in C_{\mathrm{conn}}$}{
    \ForEach{$\mathrm{site\,} i \in c$}{
        Append $R_x(i,p)$ to $\mathcal{G}$\;
        $p \gets p+1$\;
    }
}

Construct circuit $\mathcal{C}_{\mathrm{RG}}$ from $\mathcal{G}$\;
\Return{$\mathcal{C}_{\mathrm{RG}},\,p-1$}\;

\caption{Construction of the RG-based variational circuit}
\end{algorithm}

\section{Numerical details of the DMRG and VQE simulations}
\label{app:numericaldetails}
For the DMRG and VQE simulations, we use the ITensor package \cite{itensor} implemented in Julia \cite{bezanson2017julia}, which represents the state as a matrix product state. 
For the variational optimization of the tensor network state, we use the L-BFGS from the Optim package \cite{Optim} as the underlying optimizer.
For gradient calculation, we use automatic differentiation (AD) based on the algorithm given in Ref.~\cite{Guo2023}. This provides two advantages. First, the gradient is more accurate than with finite differences. Second, the computational cost is significantly reduced, as the gradient can be obtained from a single forward and backward pass, rather than requiring two function evaluations per parameter.

The variational parameters are initialized as small random numbers drawn from a normal distribution with zero mean and a standard deviation of $0.1$.
The optimization is terminated when the gradient norm falls below a tolerance of $g_{\text{tol}} = 10^{-4}$ or the maximum number of iterations, 100, is reached. 
Exact reference values are obtained via Density matrix renormalization group (DMRG) using ITensor. The Hamiltonian is represented as a matrix product operator, and the DMRG bond dimensions are increased progressively up to 800. The DMRG calculations use a cutoff of $10^{-8}$ and a maximum of 150 sweeps.
Code is available in Ref.~\cite{joch26b}.

\section{Magnitude of the gradient}
\label{app:gradient}

    A well-known challenge in variational quantum circuits is the barren plateau phenomenon \cite{McClean2018,PRXQuantum.3.010313}, which is characterized by vanishing gradients that can make optimization exponentially hard. While the mean of the gradient over random parameter configurations $\boldsymbol{\theta}$ is zero, its variance can be larger. A non-zero gradient variance is beneficial for optimization, as non-vanishing gradients are necessary to identify promising directions for parameter updates. Previous studies have shown that the gradient variance can approach zero as the circuit depth and system size increase.

\begin{figure}[htb] 
    \includegraphics[width=0.5\linewidth]{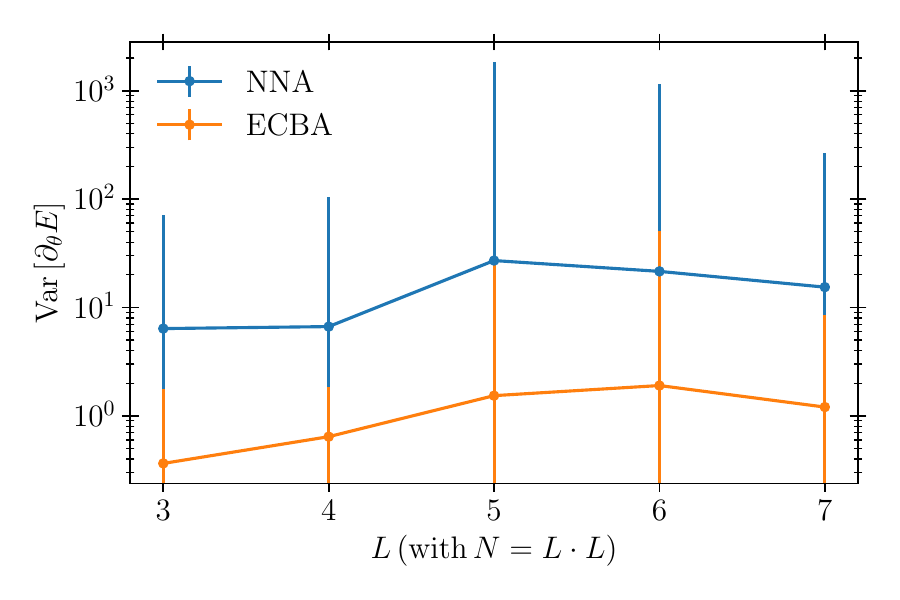}
    \vspace{-0.5em}
    \caption[]{\justifying Average variance of the gradient sampled over 1000 initial parameter sets each depending on the system size for the NNA and the ECBA. The error bars indicate the variance across the different gradient components.}
    \label{fig:gradient}
\end{figure}

    Here, we show that barren plateaus do not present a significant problem for the shallow circuits considered in our simulations. To this end, we investigate the gradient variance for both the NNA and the ECBA at different system sizes $L$, as shown in Fig.~\ref{fig:gradient}. For both ansätze, we observe that the gradient variance remains approximately constant as the system size increases.

\section{Experimental details} 
\label{app:experimental_details}
For the experiments, we use two quantum computing devices provided by IQM, namely the 20-qubit Garnet device and the 54-qubit Emerald device. Both devices use superconducting transmon qubits, which are widely used due to their relatively long coherence times and ease of control. The qubits are arranged in a square lattice and connected via tunable couplers, enabling control over their interactions. 
Both systems are calibrated to support Phased X Rotation (PRX) gates as native single-qubit gates.
These gates implement arbitrary rotations around axes in the XY plane, given by $\exp[-i\frac{\theta}{2}(\cos\phi X+\sin\phi Y)]$. The native two-qubit gate is the CZ gate \cite{abdur24}. Some relevant hardware properties are listed in Tab.~\ref{tab:iqmdevices}.
\begin{table}[htb]
\centering
    \begin{tabular}{l|c c c c}
        device  & T$_1$ & T$_2$ & PRX fidelity & CZ fidelity \\ \hline 
         Garnet  & 36.53 $\mathrm{\mu s}$ & 8.61 $\mathrm{\mu s}$ & 99.91 \% & 99.37 \% \\
        Emerald & 50.05 $\mathrm{\mu s}$ & 15.75 $\mathrm{\mu s}$ & 99.94 \% &  99.45 \%
    \end{tabular}
    \caption{\label{tab:iqmdevices} \justifying Median values of the T1 relaxation time, T2 coherence time (Ramsey), PRX gate fidelity, and CZ gate fidelity for the IQM Garnet and IQM Emerald devices \cite{abdur24}.}
\end{table}
Qubit selection and circuit transpilation are performed using the Qiskit transpile function with a square lattice embedding. This optimizes the circuits for the hardware architecture of each device without requiring SWAP gates. For the measurement, we perform readout in the $x$ and $z$ bases. These measurements are required to calculate the energy expectation value of the Hamiltonian considered in this work.
Code is available in Ref.~\cite{joch26b}.

\subsection{Error mitigation}
\label{app:em}
To improve the results from the quantum device, we employ quantum error mitigation (EM). In contrast to error correction, EM does not aim to actively correct errors in the evolution of an encoded state through measurement and feedback. Instead, it mitigates the effects of noise on specific expectation values by averaging over several modified versions of the logical circuit, combined with suitable pre- and post-processing \cite{EM,filippov2024scalabilityquantumerrormitigation}. 

To mitigate coherent errors associated with two-qubit entangling gates, we employ Pauli twirling \cite{twirling1, twirling2}. This technique involves "dressing" each two-qubit gate with randomly selected single-qubit Pauli operations while preserving its logical action. By effectively randomizing coherent noise, Pauli twirling reduces its impact and suppresses the worst-case error rate.

Readout errors are mitigated using Twirled Readout Error Extinction (TREX) \cite{trex}. In this approach, random bit-flip operations are applied immediately before measurement, effectively diagonalizing the readout error channel. The resulting measurement data are subsequently corrected through classical post-processing. This procedure improves measurement fidelity and thereby increases the reliability of the estimated observables. To obtain statistically robust results, we average over several circuit instances incorporating both TREX and Pauli twirling.

In addition, we implement dynamical decoupling \cite{dd} to suppress unwanted crosstalk, dephasing, and decoherence affecting qubits during idle periods. This is achieved by inserting sequences of single-qubit gates and their inverses into idle intervals. In our implementation, we employ a minimal sequence consisting of two Pauli-X gates during each idle sequence.

Finally, we apply digital Zero-Noise Extrapolation (ZNE) \cite{zne} to mitigate the effects of noise in the quantum circuits. This technique artificially increases the noise level by introducing logically redundant two-qubit gates, which are typically a major source of noise in superconducting quantum devices. Specifically, the entangling operations are inverted and reapplied throughout the circuit, thereby amplifying the effective noise strength from its baseline value of $\lambda = 1$ to higher odd-integer values, $\lambda\in 2\mathbb{N}+1$. The resulting expectation values are then extrapolated to the zero-noise limit, $\lambda = 0$, using an exponential noise model of the form $O(\lambda)=O_0 \, \mathrm{exp}(a\lambda)$, where $O_0$ denotes the extrapolated expectation value in the absence of noise and a is a fitting parameter. All of these error-mitigation techniques are implemented manually on top of Qiskit's \texttt{Sampler} primitive \cite{quantumcomputingqiskit}.

In practice, each observable is evaluated separately for each circuit instance, including the Pauli-twirled two-qubit gates and randomized bit-flip operations used for TREX. For the $N=7\cdot7$ system, the expectation value is estimated from 10,000 measurement shots per instance. The resulting estimates are then averaged over the 16 instances for this system, with the statistical uncertainties arising from shot noise propagated by adding the individual contributions in quadrature. The same procedure is applied for each noise-scaling factor $\lambda\in \{1, 3, 5\}$, resulting in a total of 48 circuit executions per observable for the $N=7\cdot7$ system. The error bars are obtained from a Levenberg–Marquardt fit, with the shot-noise uncertainties propagated through the fitting procedure. The resulting uncertainties are dominated by the fitting procedure rather than by shot noise.

In case of the $N=4\cdot4$ system, the expectation value is estimated from a growing number of measurement shots per instance with number of iterations, namely 2,000 shots in the first and second iteration, 5,000 shots in the third and fourth iteration, and 10,000 shots in the fifth iteration. The last three iterations are performed including error mititgation with averaging over 4 instances for each noise-scaling factor $\lambda\in \{1, 3, 5\}$, resulting in a total of 12 circuit executions per observable. The lower number of shots and instances is used because of the large number of runs needed for calculating the gradient.

\subsection{Parameter shift rule}
Automatic differentiation cannot be applied directly to quantum circuits executed on a real quantum device, as the quantum state and the operations performed on it are not explicitly accessible. Instead, we use the parameter shift rule to evaluate gradients experimentally. For a parameter $\theta$ of a gate generated by a Pauli operator, the derivative of an expectation value can be obtained as \cite{mitarai2018, schuld2019}
\begin{equation}
    \frac{\partial \langle O\rangle}{\partial \theta} = \frac{1}{2} \left[ \langle O\rangle_{\theta+\frac{\pi}{2}} -\langle O\rangle_{\theta-\frac{\pi}{2}} \right] \, .
\end{equation}
Thus, the gradient is determined from two circuit evaluations with shifted parameter values. Unlike finite difference methods, the parameter shift rule provides an exact derivative for gates generated by Pauli operators.

\section{Convergence behavior} 
\label{app:convergence}

\begin{figure}[t] 
    \includegraphics[width=0.5\linewidth]{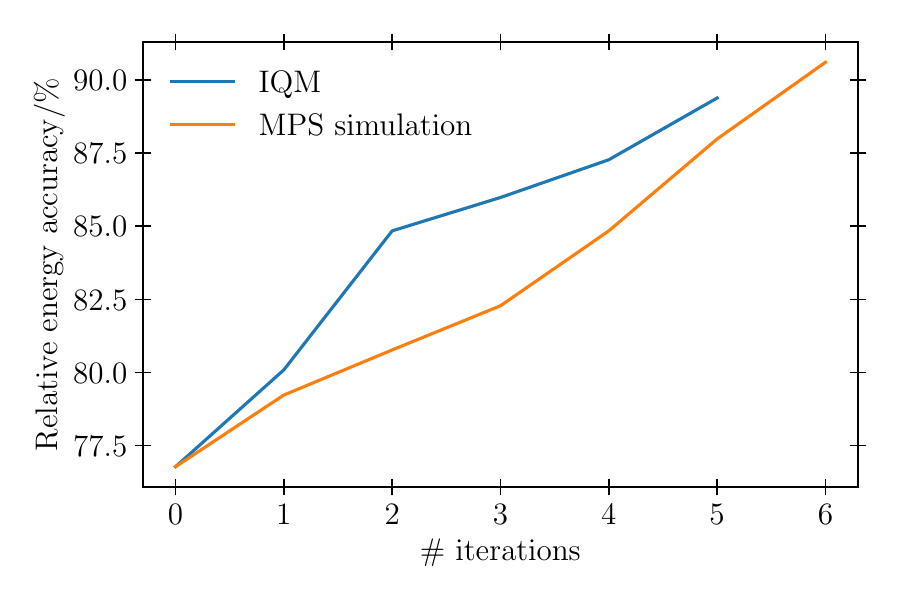}
    \vspace{-0.5em}
    \caption[]{\justifying Convergence behavior of the $L=4$ system starting from the RG solution (iteration 0) for the IQM calculation compared to the simulation with ITensor. A bond dimension of 20 is used in the latter case.}
    \label{fig:convergence}
\end{figure}

For the system with $L=4$, we perform a total of five iterations on the IQM device. Figure~\ref{fig:convergence} compares the relative energy accuracy obtained after each iteration with the corresponding results from MPS simulations. We observe good agreement between the two curves, indicating that the experimental results closely follow the expected behavior.

However, carrying out further iterations on the IQM device requires an increasing number of measurement shots and stronger EM, as discussed in Appendix~\ref{app:em}. This leads to a progressively longer runtime, which limits our calculations to five iterations. Nevertheless, we expect that additional iterations would be possible with a sufficiently large number of shots and sufficiently effective EM, as long as the errors remain smaller than the gradient we aim to resolve.

\end{appendix}

\end{document}